\documentclass[aps,amsmath,superscriptaddress,floatfix,nobalancelastpage,pra]{revtex4}

\usepackage{amsmath}
\usepackage{mathtools}
\usepackage{amssymb}
\usepackage{amsthm}
\usepackage{graphicx}
\usepackage{color}
\usepackage{hyperref}

\newcommand{\RR}{\mathbb{R}}

\newcommand{\paren}[1]{\left( #1 \right)}

\newcommand{\wh}[1]{\widehat{#1}}
\newcommand{\wt}[1]{\widetilde{#1}}
\newcommand{\wb}[1]{\overline{#1}}
\renewcommand{\Re}{\operatorname{Re}}
\renewcommand{\Im}{\operatorname{Im}}

\DeclareMathOperator\erf{erf}
\newcommand{\abs}[1]{\lvert#1\rvert}

\newcommand{\Schrod}{Schr{\"o}dinger }

\newcommand{\vks}{V_{\text{KS}}}
\newcommand{\vecA}{A}
\newcommand{\Vpsi}{\paren{V\psi}}
\newcommand{\Vpsij}{\paren{V\psi_j}}
\newcommand{\psihat}{\wh{\psi}}
\newcommand{\Vpsihat}{\wh{\paren{V\psi}}}
\newcommand{\psijhat}{\wh{\psi_j}}
\newcommand{\psijzhat}{\wh{\psi_{j,0}}}
\newcommand{\Vpsijhat}{\wh{\paren{V\psi_j}}}
\newcommand{\psijtil}{\wt{\psi_j}}
\newcommand{\psijztil}{\wt{\psi_{j,0}}}

\newcommand{\CCQ}{Center for Computational Quantum Physics, Flatiron
Institute, 162 5th Ave., New York, NY 10010, USA}
\newcommand{\CCM}{Center for Computational Mathematics, Flatiron
Institute, 162 5th Ave., New York, NY 10010, USA}
\newcommand{\courant}{Courant Institute of Mathematical Sciences, New
York University, 251 Mercer St., New York, NY 10012, USA}
\newcommand{\MPI}{Max Planck Institute for the Structure and Dynamics of Matter and Center for Free Electron Laser Science, 22761 Hamburg, Germany}
\newcommand{\UNIPA}{Dipartimento di Fisica e Chimica - Emilio Segr\`e, Universit\`a degli Studi di Palermo, Via Archirafi 36, I-90123, Palermo, Italy}

\begin{document}

\title{Eliminating artificial boundary conditions in time-dependent
density functional theory using Fourier contour deformation}

\author{Jason Kaye}
\email{jkaye@flatironinstitute.org}
\affiliation{\CCM}
\affiliation{\CCQ}

\author{Alex Barnett}
\affiliation{\CCM}

\author{Leslie Greengard}
\affiliation{\CCM}
\affiliation{\courant}

\author{Umberto De Giovannini}
\affiliation{\UNIPA}
\affiliation{\MPI}

\author{Angel Rubio}
\affiliation{\CCQ}
\affiliation{\MPI}

\begin{abstract}

  We present an efficient method for propagating the time-dependent Kohn-Sham
  equations in free space, based on the recently introduced
  Fourier contour deformation (FCD) approach. For potentials which are
  constant outside a bounded domain, FCD yields a high-order accurate numerical solution
  of the time-dependent \Schrod equation directly in free space, without
  the need for artificial boundary conditions. Of the many existing
  artificial boundary condition schemes, FCD is most similar to an exact
  nonlocal transparent boundary condition, but it works directly
  on Cartesian grids in any dimension, and runs on top of the fast
  Fourier transform rather than fast algorithms for the application of
  nonlocal history integral operators. We adapt FCD to time-dependent
  density functional theory (TDDFT), and describe a simple algorithm to smoothly
  and automatically truncate long-range Coulomb-like potentials to a
  time-dependent constant outside of a bounded domain of interest, so
  that FCD can be used. This approach eliminates errors originating from
  the use of artificial boundary conditions, leaving only the error of
  the potential truncation, which is controlled and can be systematically
  reduced. The method enables accurate simulations of ultrastrong
  nonlinear electronic processes in molecular complexes in which the
  inteference between bound and continuum states is of paramount
  importance. 
  We demonstrate results for many-electron TDDFT calculations of
  absorption and strong field photoelectron spectra
  for one and two-dimensional models, and observe a significant reduction in
  the size of the computational domain required to achieve high quality
  results, as compared with the popular method of complex absorbing
  potentials.

\end{abstract}

\maketitle

\date{}

\section{Introduction}

Intense infrared and deep-infrared
laser spectroscopy provide structural information
on atomic and molecular gases which enables the inspection of
ultra-fast and ultra-small-scale physics~\cite{Ivanov.2021}.
A variety of experimental techniques, such
as high harmonic generation, photoelectron spectroscopy, and
photo-fragmentation, have been used to study the nonlinear dynamics
induced by strong applied fields. Experiments like these have addressed
a number of fundamental questions involving, for example, the
characteristic time scales of correlations~\cite{Mansson.2021}, the
electronic and structural changes which molecules undergo following a
chemical reaction, and sub-cycle electron
dynamics~\cite{Trabattoni.2020}. As a consequence of their small mass, electrons 
are the first to respond to strong fields, and the development of 
theoretical and numerical methods to simulate nonlinear many-electron 
dynamics has underpinned tremendous progress in this field.

Simulating strong field experiments with increasingly large and complex
polyatomic molecules requires the use of models capable of capturing the
dynamics of many electrons interacting with the field, and accounting for
interference between bound and continuum states
\cite{Smirnova.2009,Boguslavskiy.2012}. Ab-initio
approaches like time-dependent density functional theory (TDDFT), which
involve no empirically fitted parameters, have become the leading approach
in many-electron simulations. TDDFT reformulates
the many-electron time-dependent \Schrod equation (TDSE) as a
system of nonlinear single-electron equations, coupled only through
a dependence of the electronic potential on the total electron density.
Electron-electron interactions are modeled using an exchange and
correlation term in the potential, yielding an approximation of the
full $N$-electron dynamics at the tractable cost of $N$
single-electron simulations \cite{Runge.1984,Marques.2011,Ullrich.2012}.

Numerical methods used in the electronic structure community to simulate photo-chemical processes involving
ionization or scattering typically solve the underlying equations on a
bounded domain of interest, and discard the outgoing part of the
wavefunction. Indeed, for large scale simulations at long propagation
times, it is computationally infeasible to track the full wavefunction
on its entire domain of support. Instead, the true free space dynamics
are mimicked using an artificial boundary scheme, which may take
the form of an approximate or exact outgoing boundary condition, or an
absorbing boundary layer. Even for a single electron obeying the linear
time-dependent \Schrod equation (TDSE), designing effective artificial
boundaries is a
challenging task, and there is an extensive literature on the
subject. The available methods include complex absorbing
potentials (CAPs) \cite{Navarro.2004}, mask functions (MFs)
\cite{Krause.19922x,DeGiovannini.2012}, exterior complex scaling (ECS)
\cite{Simon.1979m99,Balslev.1971,Moiseyev.2011}, perfectly matched
layers (PMLs) \cite{johnson21,antoine17,nissen11,mennemann14,antoine20}, local artificial
boundary conditions (ABCs) \cite{engquist77,antoine01,antoine04,Antoine.2008}, and exact nonlocal transparent boundary
conditions (TBCs)
\cite{baskakov91,jiang04,jiang08,schadle02,lubich02,schadle06,han04,han07,ermolaev99,feshchenko11,feshchenko13,feshchenko17,kaye18,kaye20,ji18,arnold03,arnold12}. We refer to Refs.
\cite{Antoine.2008,antoine17} for useful
introductions to the literature, and more extensive collections of references.

The use of artificial boundaries has become popular in numerical quantum
chemistry, particularly in real time time-dependent electronic structure theory
\cite{Li.2020l4j}. This includes, beyond TDDFT with hybrids~\cite{Isborn.2007},
correlated wavefunction methods such as multi-configurational
self-consistent-field (MCSCF)~\cite{Sato.2013}, time-dependent configuration
interaction (TD-CI)~\cite{Krause.2005}, and coupled cluster (CC)~\cite{Nascimento.2016}.
These real time techniques are essential for simulating the non-perturbative electronic response 
to electric fields in the strong-field regime, including phenomena such as
charge redistribution, multi-photon absorption, high-harmonic generation, and
strong-field ionization, in addition to photoelectron spectroscopy, which is the
focus of this paper~\cite{Schlegel.2007,Smith.2005,Lopata.2013,Luppi.2012,Sissay.2016,Zhu.2022,Krause.2014,Krause.2014y3d,Krause.2015,Suzuki.2003,Suzuki.2004}.

In large scale TDDFT calculations, simple local methods such as CAPs are usually favored.
A CAP is a purely imaginary potential added to the system Hamiltonian,
which becomes nonzero outside of a subdomain of interest. This leads to
damping of the wavefunction within the support of the CAP, which we
refer to as an absorbing layer.
CAPs are not restricted to grid-based 
numerical methods and have been applied to atomic-centered basis calculations 
\cite{Luppi.2012,Sissay.2016,Zhu.2022} even in combination with TD-CI \cite{Krause.2014,Krause.2014y3d,Krause.2015}.
For good performance---that is, to successfully approximate the true
free space wavefunction without an excessively large absorbing layer, which
itself must be discretized by a grid---CAPs require significant tuning of their functional form,
the width of the absorbing layer, and the momentum components targeted
for damping. We note that the popular MF method, 
which imposes an explicit damping of the wavefunction at each time
step, is equivalent to a particular choice of CAP to first order accuracy
in the time step size \cite{DeGiovannini.2015}.

We also mention ECS \cite{Simon.1979m99}, a generalization of the older
complex scaling method \cite{Balslev.1971,Moiseyev.2011}, which uses an
analytical continuation of the Hamiltonian obtained by rotating the
spatial coordinates into the complex plane outside of a subdomain of
interest. This complex rotation results in exponential damping of the
outgoing part of the wavefunction. Although the original goal of ECS
was to investigate the energy and lifetimes of shape resonances as a
static problem~\cite{Whitenack.2011ce1vm,Larsen.2013}, it has more
recently been applied to the implementation of artificial boundaries for the TDSE, and is now
considered part of the state of the art in strong field simulations
\cite{Schneider.2020}.
In particular, in infinite range ECS (irECS), an unbounded element is added to a finite element
discretization of the TDSE in order to represent outgoing waves damped
by an ECS transformation \cite{Scrinzi.2010}.
As for CAPs, irECS requires tuning parameters which control the damping
of the wavefunction. In addition, despite progress in implementing
variants of irECS for grid-based methods like finite differences
\cite{Weinmuller.2017}, applications so far have been limited to highly
symmetric systems, like atoms and dimers, amenable to the use of polar
or spherical coordinate grids
\cite{Telnov.2013,Dujardin.2014,Orimo.2018,Zhu.2020,Scrinzi.2022}.
These grids enable the straightforward discretization of exterior
domains by an element which is unbounded in the radial coordinate, but
are inefficient for the more general molecular geometries often encountered in
large scale TDDFT calculations.

Recently, a new method was proposed by some of the authors which eliminates the need for
artificial boundaries
for the linear TDSE in the case that the potential is compactly
supported, or more generally, constant outside of a bounded domain
\cite{kaye22_tdse}. This method, which we refer to as Fourier contour
deformation (FCD), represents the wavefunction as a
Fourier series of complex-frequency modes, obtained from a contour
deformation of the spatial Fourier variable. Such a representation
correctly captures the outgoing components of the wavefunction, to a
user-controllable accuracy, without imposing artificial boundaries of any kind.
Waves may even leave the computational domain and return later under the
influence of an applied field, without any loss of accuracy, offering a contrast
with methods like CAPs and MFs, which damp outgoing waves, and
irECS, which both damps outgoing waves and amplifies incoming waves.
The method has several other advantages: it is spectrally accurate in
space, can be coupled to high-order accurate time stepping methods,
operates naturally on Cartesian grids in any dimension, and analytically
includes the influence of applied fields.

The purpose of this paper is to adapt the FCD method to realistic many-electron
light-matter simulations using TDDFT with regular, real space grids. This is accomplished by first
applying a smooth and controlled 
truncation of the long range Coulomb-like potential
appearing in TDDFT to a potential which is spatially constant outside of
a bounded domain of interest, and then using FCD, suitably
modified for the nonlinear Kohn-Sham equations. We then demonstrate,
using an implementation in the Octopus code for TDDFT
\cite{Tancogne-Dejean.2020}, that
such a truncation can be made without significantly affecting
important physical observables, and show that our
approach therefore allows for the use of smaller computational domain
sizes than those obtained on a Cartesian grid using CAPs.

Our approach differs in an important way from that of most existing
artificial boundary schemes. All methods, including our
own, must in effect make a truncation or modification of the electrostatic potential
outside a subdomain of interest,
though only some do so explicitly. Indeed, no method can include the
effect of a generic potential on the solution outside of the
computational domain. However, this is the only approximation our method
makes---specifically, we make an explicit truncation of the potential to a
time-dependent constant outside the subdomain of interest, and the resulting
equations are solved exactly, up to controlled discretization errors,
with no further artificial boundary scheme. By contrast, even for potentials which are
constant or zero outside of a bounded domain, CAPs, ECS, PMLs, and local
ABCs all require a further modification of the underlying equations in
order to avoid boundary reflections, necessarily introducing additional
errors which can be difficult to quantify.

Exact nonlocal TBC methods
share this feature with the FCD method, and can be considered its close
cousins. These methods have
not, however, gained widespread adoption in the context of TDDFT,
because of several practical barriers. Exact nonlocal TBCs must be
evaluated using specialized fast algorithms in order to avoid a rapid growth of the
computational cost with the propagation time, and such algorithms, like
ECS, typically rely on the use of polar or spherical coordinate grids
\cite{jiang08,han04,han07,arnold12}.
Cartesian grid-based methods which have been proposed either lack fast algorithms,
leaving them computationally intractable for large scale problems
\cite{feshchenko11,feshchenko13,ji18}, or rely on specific low-order
spatial discretizations \cite{arnold03,arnold12,mennemann14,ji18}.
An exception, Ref. \cite[Chapter 2]{kaye20}, is significantly more complicated than
the approach described here, with no obvious advantages.

This paper is organized as follows. In Section \ref{sec:tdkse}, we
briefly review TDDFT and the time-dependent Kohn-Sham equations. In
Section \ref{sec:method}, we review the FCD method, and describe our potential truncation scheme. In
Section \ref{sec:results}, we compare the method with CAPs in
simulations of absorption and photoemission spectroscopy. We give
concluding remarks and outline future directions of research in Section
\ref{sec:conclusion}. Appendix \ref{app:propagator} suggests a particular time
discretization, and gives other implementation details.

Atomic units (a.u.), with $e=\hbar=m_e=1$, are used
throughout, unless otherwise specified. 

\section{Time-dependent Kohn-Sham equations} \label{sec:tdkse}

TDDFT provides a systematic and computationally tractable method of
simulating the many-body TDSE using a collection of single-particle
nonlinear TDSEs, called the time-dependent Kohn-Sham equations (TDKSE)~\cite{Runge.1984,
Marques.2011}. The solutions
$\psi_j(x,t)$ are called the Kohn-Sham orbitals, and represent a
collection of fictitious, non-interacting electronic wavefunctions whose
Slater determinant reproduces the time-dependent many-body electron
density $\rho(x,t)$. The
equations are nonlinearly coupled through a shared potential $\vks[\rho](x,t)$, called the
Kohn-Sham potential, which depends only on the density
of the Kohn-Sham electrons, and not on the individual
orbitals. The TDKSE for $N$ electrons in an
electromagnetic field, as well as the definitions of $\vks$ and $\rho$, are
given as follows:
\begin{equation} \label{eq:tdkse}
  \begin{aligned}
    i \partial_t \psi_j(x,t) &= \frac{1}{2}\paren{-i\nabla -
    \vecA(t)}^2 \psi_j(x,t) + \vks[\rho](x,t) \psi_j(x,t) \\
    \psi_j(x,0) &= \psi_{j,0}(x) \\
    \vks[\rho](x,t) &= v_{\rm ion}(x,t)+v_{\rm H}[\rho](x,t)+v_{\rm
    xc}[\rho](x,t)\\
    \rho(x,t) &= \sum_{n=1}^N \abs{\psi_j(x,t)}^2.
  \end{aligned}
\end{equation}
Here, the coupling with the external electromagnetic field is described in
the velocity gauge and the long-wavelength approximation by
means of a spatially uniform vector potential $\vecA(t)$, which induces
an electric field $E(t) = -\frac{dA(t)}{dt}$. A factor $1/c$, with $c$ the
speed of light, has been absorbed into $A(t)$. The Kohn-Sham potential
$\vks[\rho](x,t)$ is divided into several parts. $v_{\rm ion}(x,t)$ is the
potential due to the atomic nuclei. $v_{\rm H}[\rho](x,t)$, the Hartree
potential, is the classical electrostatic potential due to a charge
density $\rho$, and accounts for part of the electron-electron interaction
energy. $v_{\rm xc}[\rho](x,t)$, the exchange and correlation potential,
accounts for the remainder of the electron-electron interaction.
Although it can be shown that there exists a choice of $v_{\rm
xc}[\rho](x,t)$ such that the exact many-body electron density can be recovered
from the Kohn-Sham orbitals, no systematic method exists of constructing
it. However, successful proposed approximations of this term, valid for
large classes of physical systems, have made TDDFT a leading framework
for many-electron simulations.

We note that although there exist a variety of proposals for more accurate exchange and
correlation potentials which depend not only on the electron
density $\rho$, but also on the individual Kohn-Sham orbitals $\psi_j$
and/or currents,
for simplicity we restrict our discussion here to the classical case.
There is no significant barrier to extending our method to the more
general setting. In addition, we focus on the case of closed-shell systems
in which all orbitals are doubly occupied, but our
method also applies to more sophisticated formulations involving, for
instance, open shells or non-collinear spins \cite{vonBarth:1972eq}.

\section{Eliminating artificial boundary conditions using Fourier
contour deformation} \label{sec:method}

We present a method to solve the TDKSE \eqref{eq:tdkse} in free space,
$x \in \RR^d$ with $d$ the spatial dimension, which avoids the problem of artificial boundary
conditions. Our explanation proceeds in two steps. First, we describe the
Fourier contour deformation (FCD) method \cite{kaye22_tdse} for \eqref{eq:tdkse} in the case of a Kohn-Sham potential which
is constant in space, $\vks[\rho](x,t) = v(t)$ for a smooth function $v$,
outside of a bounded computational domain $[-L,L]^d$. This method solves
the equation correctly, up to discretization errors, without requiring any artificial
boundary conditions, and returns the solution on $[-L,L]^d$.
Second, we propose a simple method for truncating the long-range
potentials encountered in many practical calculations to a
time-dependent constant $v(t)$.

\subsection{FCD for potentials which are constant outside of a bounded domain}
\label{sec:fcd}

Our discussion follows Ref. \cite{kaye22_tdse}, which introduced the FCD method for the
linear TDSE with compactly supported potentials. Here, we expand the
discussion to include nonlinear potentials of the form in
\eqref{eq:tdkse} which are spatially constant
outside of $[-L,L]^d$. We give a brief review, and
refer the reader to the original reference for a detailed description
and analysis. We assume throughout that the initial Kohn-Sham orbitals
$\psi_{j,0}(x)$ are supported in $[-L,L]^d$, but this assumption can be
lifted in certain cases.

To simplify notation, we define $V[\rho](x,t) \equiv \vks[\rho](x,t) +
\frac12 \abs{\vecA(t)}^2$, with $\abs{\cdot}$ denoting the Euclidean
norm, so that the first equation in \eqref{eq:tdkse}
becomes
\begin{equation} \label{eq:tdkse2}
  i \partial_t \psi_j(x,t) = \paren{-\frac12 \nabla^2 + i \vecA(t) \cdot
  \nabla + V[\rho](x,t)} \psi_j(x,t).
\end{equation}
We note that if $V[\rho](x,t) = v(t)$ outside of $[-L,L]^d$, for a
smooth function $v(t)$, then up
to a gauge transformation, we
can assume $v(t) = 0$, i.e. that $V[\rho]$ is compactly supported. More
specifically, we can simply solve
\eqref{eq:tdkse2} with $V[\rho](x,t)$ replaced by $V[\rho](x,t) -
v(t)$, and recover the original solution as $e^{-i \int_0^t v(s) \, ds}
\psi_j(x,t)$. We therefore assume in the following discussion that
$V[\rho]$ is compactly supported. For
brevity of notation, we consider the solution for a single state
$\psi_j$, and suppress the state index $j$. We also sometimes suppress
the dependence of $V$ on $\rho$, writing simply $V$, and denote the
product $V[\rho](x,t) \psi(x,t)$ by $\Vpsi(x,t)$.

Taking the Fourier transform
of \eqref{eq:tdkse2} gives
\begin{equation} \label{eq:tdsefourier}
  \begin{aligned}
  i \partial_t \psihat(\xi,t) &= \paren{\frac12 \abs{\xi}^2 - \vecA(t)
  \cdot \xi} \psihat(\xi,t) + \Vpsihat(\xi,t) \\
  \psihat(\xi,0) &= \psihat_0(\xi),
  \end{aligned}
\end{equation}
where we have defined the Fourier transform as
\begin{equation}\label{eq:ft}
  \wh{f}(\xi) \equiv \int_{\RR^d} e^{-i \xi \cdot x} f(x) \, dx.
\end{equation}

\begin{figure}
  \centering
  \includegraphics[width=0.35\textwidth]{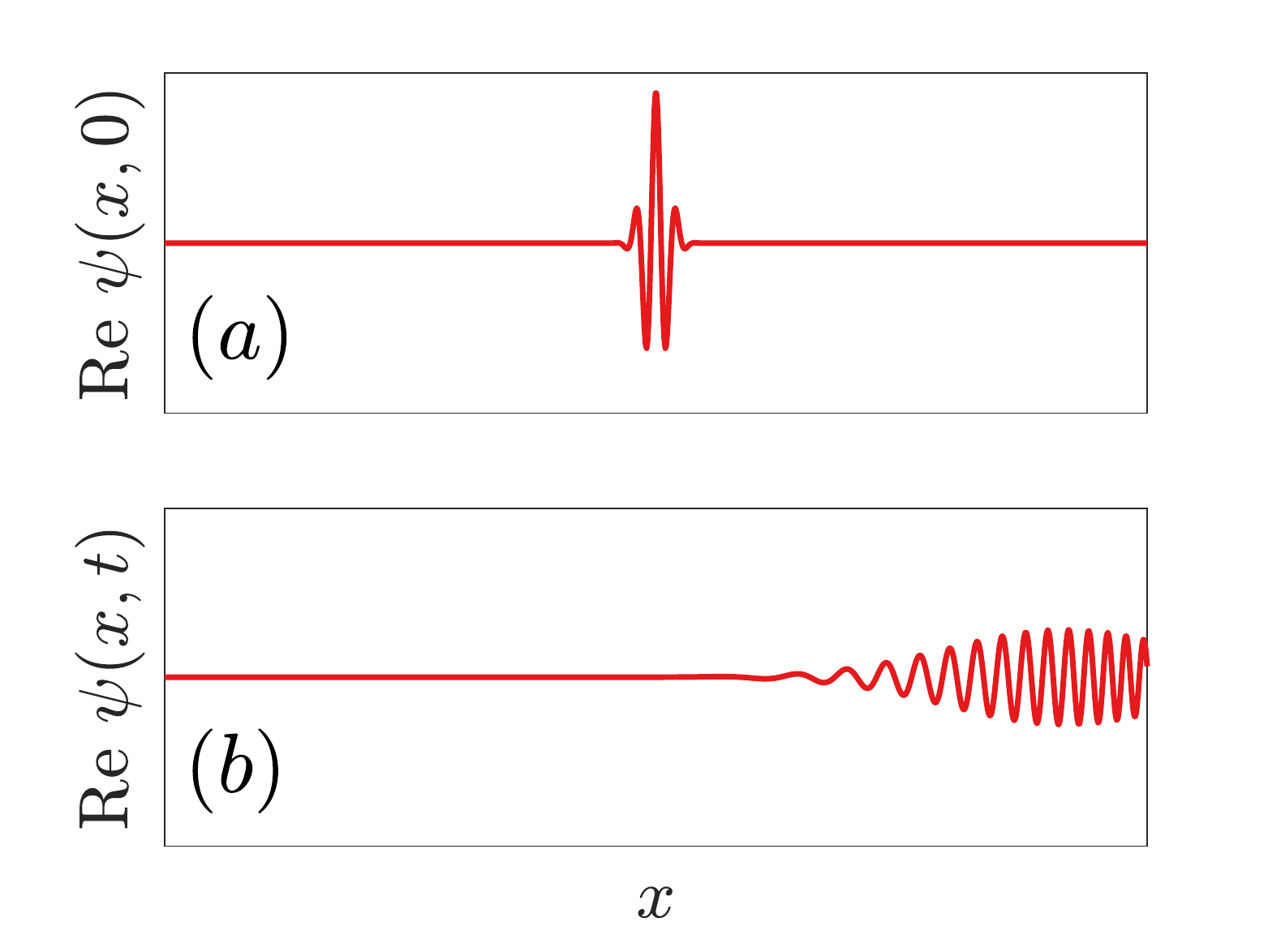}
  \includegraphics[width=0.35\textwidth]{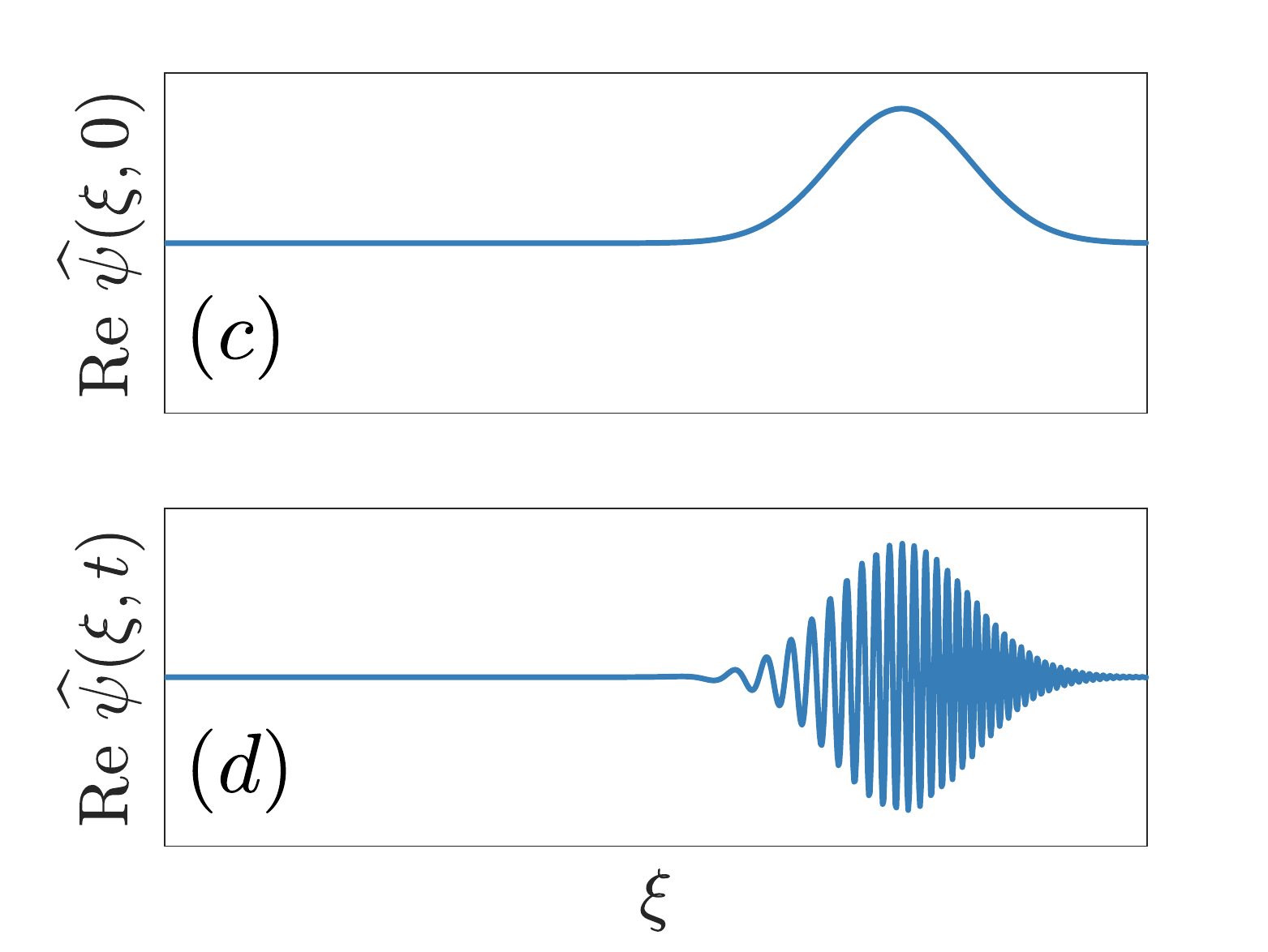}
  \includegraphics[width=0.35\textwidth]{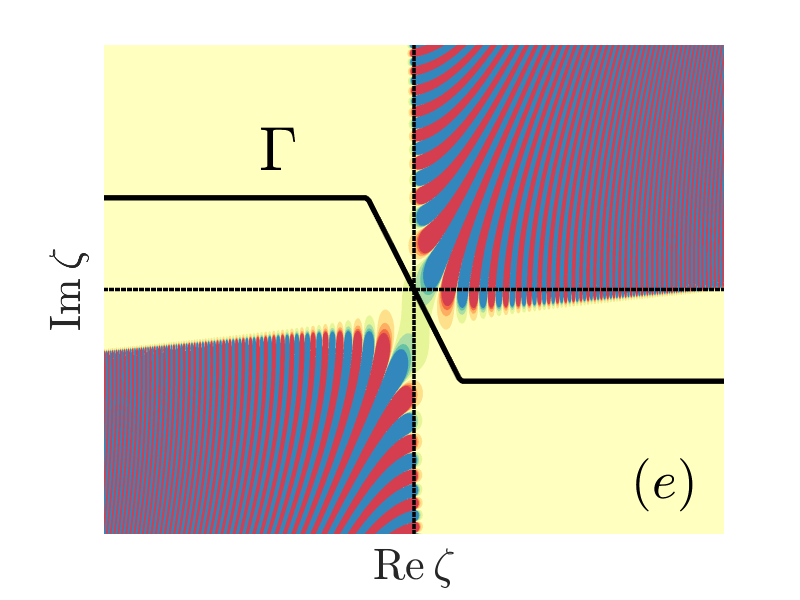}
  \includegraphics[width=0.35\textwidth]{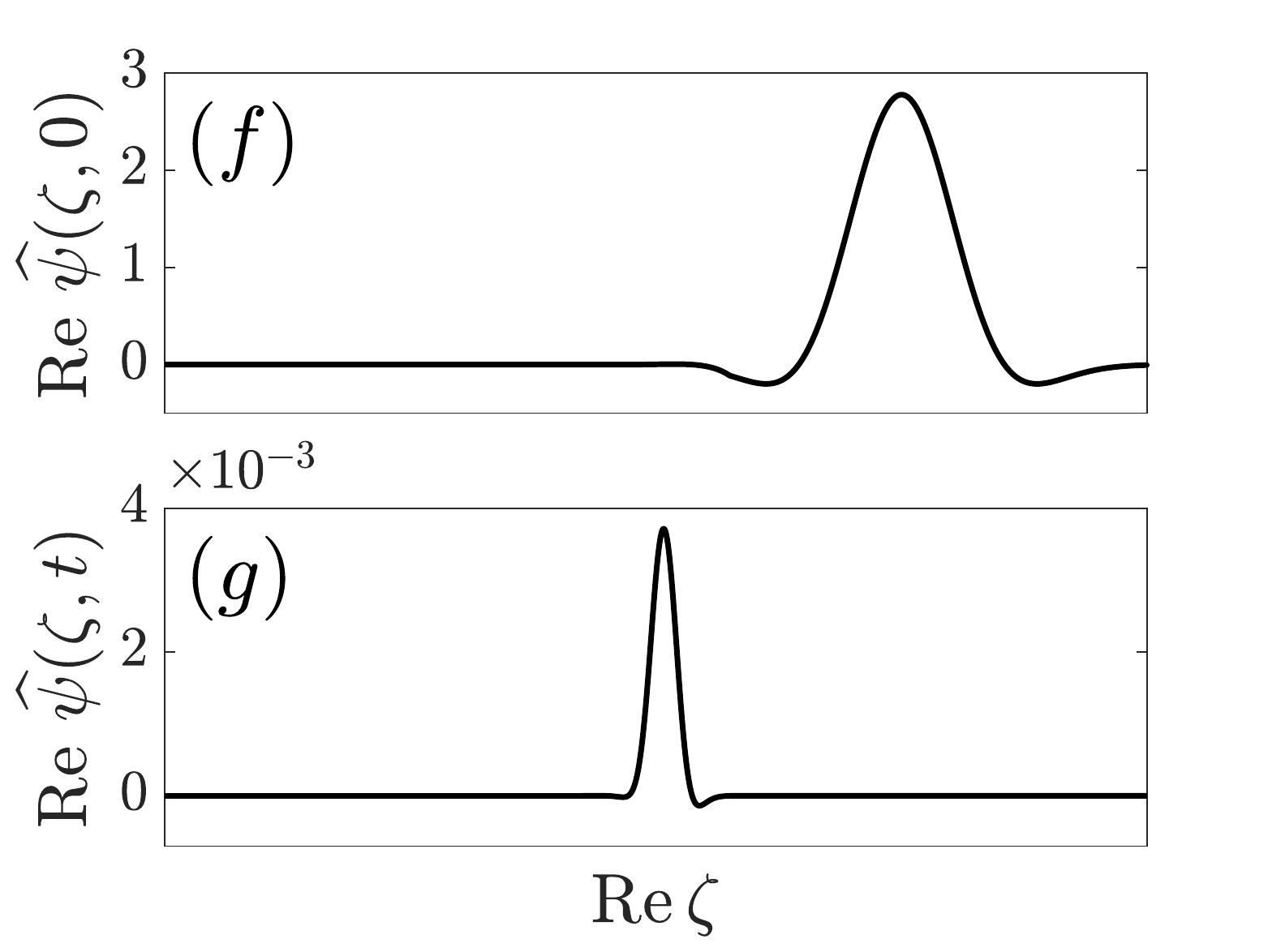}
  \caption{(a,b) A Gaussian wavepacket $\psi(x,t)$ with positive
  momentum, initially centered at the origin, travels rightward and
  spreads over time. (c,d) As a result, its Fourier transform
  $\psihat(\xi,t)$ becomes more and more oscillatory; the number of
  oscillations in a given momentum window is proportional to $t$. (e) In
  the complex $\zeta$ plane, $\psihat(\zeta,t)$ has oscillatory regions,
  including on the positive real axis, and non-oscillatory regions. The
  contour $\Gamma$ avoids the oscillatory regions. Here, $\Re
  \psihat(\zeta,t)$ is shown on a color plot with blue indicating
  positive values, and red indicating negative values. (f,g) Along the contour $\zeta \in
  \Gamma$, the oscillations in $\psihat(\zeta,t)$ are damped
  exponentially, resulting in a much smoother function than
  $\psihat(\xi,t)$. Nevertheless, the wavefunction $\psi(x,t)$, restricted
  to the domain of interest depicted in (b), can be recovered with
  controllable accuracy from $\psihat(\zeta,t)$, $\zeta \in \Gamma$.
  Note that the values in (f) are smaller than those in (e) because
  $\psihat(\zeta,t)$ is exponentially damped along $\zeta \in \Gamma$ as
  $t$ increases.}
  \label{fig:contour}
\end{figure}

One could then attempt to solve \eqref{eq:tdsefourier} by a
pseudospectral method, proceeding as follows: 
\begin{enumerate}
  \item Given $\psihat(\xi,t-\Delta t)$ and $\Vpsihat(\xi,t - \Delta
    t)$, take a time step of size $\Delta t$ by solving \eqref{eq:tdsefourier} to obtain $\psihat(\xi,t)$.
  \item Apply the inverse Fourier transform to obtain $\psi(x,t)$.
  \item Compute $\Vpsi(x,t)$, use it to compute $\Vpsihat(\xi,t)$, and repeat.
\end{enumerate}
Since $V$ is compactly supported, such a
method already appears to eliminate the problem of artificial boundary
conditions. Indeed, we are only ever required to compute the Fourier transform of
$\Vpsi(x,t)$, which is compactly supported, and $\psi(x,t)$ need
never be evaluated outside of this support. The method does not require
specifying computational domain boundaries or boundary conditions.
However, as a result of the spreading of $\psi(x,t)$ over time, the
number of oscillations in $\psihat(\xi,t)$ tends to grow rapidly. Figs.
\ref{fig:contour}(a)-(d) give a demonstration in the case of a
Gaussian wavepacket. This limits the method to short propagation times.
In particular, it can be shown that in general there is no significant
difference between the number of degrees of freedom required to
represent $\psi(x,t)$ on its full numerical support and that required to
resolve $\psihat(\xi,t)$ \cite{kaye22_tdse}.

To address this problem, we begin by writing $\psi(x,t)$ in terms of its inverse Fourier transform:
\begin{equation}\label{eq:psiift}
  \psi(x,t) = \frac{1}{(2 \pi)^d} \int_{\RR^d} e^{i \xi \cdot x}
  \psihat(\xi,t) \,  d \xi.
\end{equation}
In Ref. \cite{kaye22_tdse}, it is shown that $\psihat(\xi,t)$ is analytic in the entire
complex plane, and the integrals in
\eqref{eq:psiift} can therefore be deformed to a suitable contour $\Gamma$:
\begin{equation}\label{eq:psizift}
  \psi(x,t) = \frac{1}{(2 \pi)^d} \int_{\Gamma^d} e^{i \zeta \cdot x}
  \psihat(\zeta,t) \, d \zeta.
\end{equation}
It can be shown that along the contour $\Gamma$ depicted in Figure \ref{fig:contour}(e), the
oscillations in $\psihat$ are damped exponentially, with the rate of
damping proportional to the rate of oscillation. As a result, the number of degrees of freedom required to
resolve $\psihat(\zeta,t)$, for $\zeta \in \Gamma^d$, grows only
logarithmically in time. Figs. \ref{fig:contour}(f,g) illustrate the
smoothness of $\psihat(\zeta,t)$ on $\Gamma$ for the Gaussian wavepacket
example. Thus, we can simply propagate \eqref{eq:tdsefourier} by the
method described above, with $\xi$
replaced by the complex contour variable $\zeta \in \Gamma^d$:
\begin{equation} \label{eq:tdsegamma}
  \begin{aligned}
      i \partial_t \psihat(\zeta,t) &= \paren{\frac12 \zeta\cdot\zeta - \vecA(t) \cdot \zeta}
      \psihat(\zeta,t) + \Vpsihat(\zeta,t) \\
      \psihat(\zeta,0) &= \psihat_0(\zeta).
  \end{aligned}
\end{equation}
Note that we have written $\zeta \cdot \zeta = \sum_{l=1}^d
\zeta_l^2$ to distinguish this
quantity from $\abs{\zeta}^2 =  \sum_{l=1}^d \abs{\zeta_l}^2$.
We emphasize that the shape of the contour $\Gamma$ depicted in Figure
\ref{fig:contour}(e) need not be hand-picked given a particular problem.
Rather, this shape, which consists of two horizontal components and a
diagonal component passing through the origin, is always used. The only
parameter associated with $\Gamma$ is the distance from the real axis of its
horizontal components, which is chosen based on a user-specified error
tolerance and the ``quiver radius'' of the applied field, defined as
$\max_t \abs{\int_0^t A(s) \, ds}$.

Our high-order accurate implementation of the FCD method for TDDFT is
described in detail in Appendix \ref{app:propagator}. It involves a
high-order discretization of \eqref{eq:psizift}, which leads to an
approximation of $\psi(x,t)$, valid in $[-L,L]^d$, as a Fourier series
with complex-frequency modes:
\[\psi(x,t) \approx \frac{1}{(2\pi)^d} \sum_{k=1}^N e^{i \zeta_k \cdot x}
\psihat(\zeta_k,t) w_k.\]
Here, the $\zeta_k$ are discretization nodes on $\Gamma^d$, and $w_k$
are quadrature weights, which can be absorbed into $\psihat(\zeta_k,t)$
to obtain Fourier series coefficients. This gives an intuitive
explanation of the FCD method: whereas for periodic problems, a Fourier
series of integer modes gives an accurate representation of the
wavefunction on its domain, for free space problems, a Fourier series of
complex-frequency modes gives an accurate representation on $[-L,L]^d$.

\subsection{Truncation of long-range potentials}

Now that we have described a method of solving \eqref{eq:tdkse} for
potentials $V$ which are spatially constant outside of a bounded
computational domain, we must show how the potentials encountered in
practical TDDFT calculations, which typically exhibit Coulomb-like
decay, can be put into this form.
We propose a simple method for smoothly rolling the potential off
to a time-dependent constant $v(t)$ outside the sphere $S_L^{d-1}$ of radius $L$
centered at the origin.
We take $v(t)$ to be the average of the unmodified potential over this sphere:
\begin{equation} \label{eq:vavg}
  v(t) = \frac{\Gamma\paren{d/2}}{2 \pi^{d/2} L^{d-1}} \int_{S_L^{d-1}}
  V[\rho](x,t) \, dS(x).
\end{equation}
For $d = 1$, we define
\[v(t) = \frac12 \paren{V[\rho](-L,t) + V[\rho](L,t)}.\]

\begin{figure}
  \centering
  \includegraphics[width=0.49\textwidth]{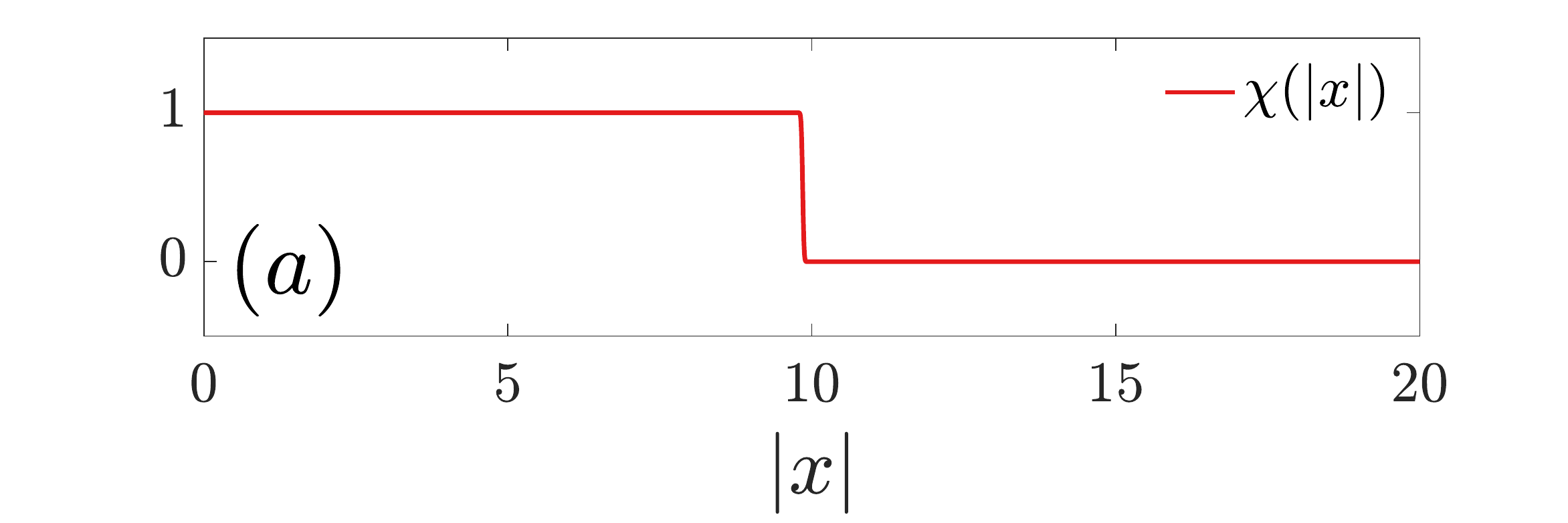}
  \includegraphics[width=0.49\textwidth]{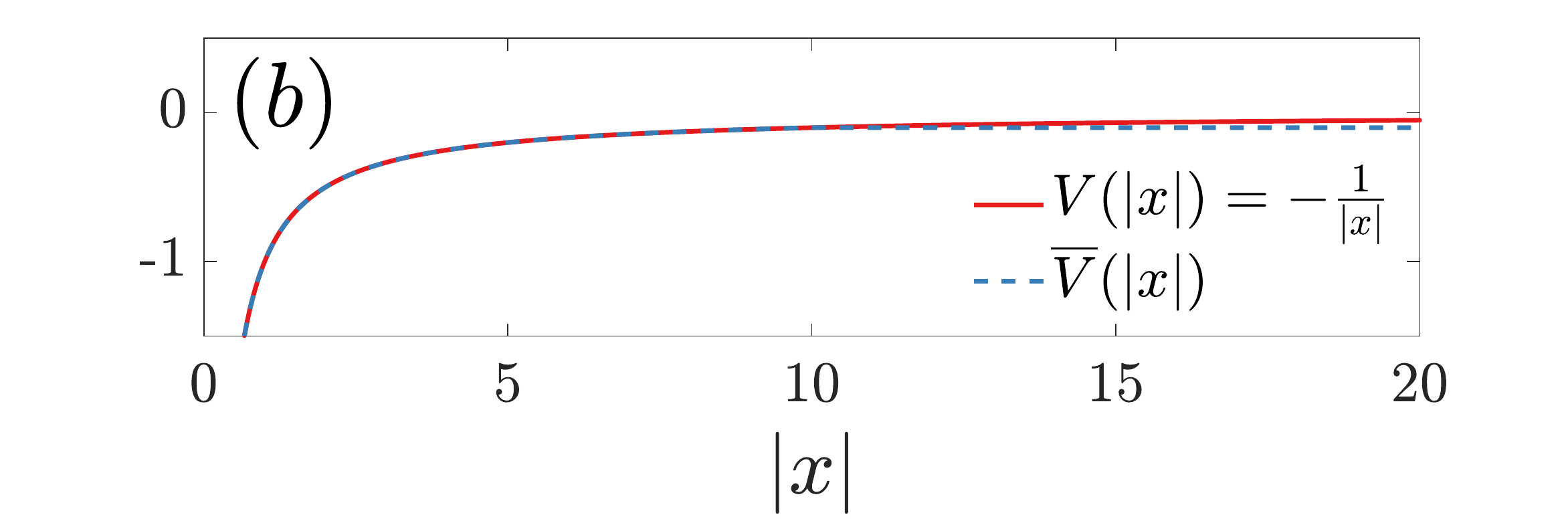}
  \caption{(a) The bump function $\chi(\abs{x})$ with $L = 10$ and
  $\sigma = 0.03L$. (b) The Coulomb potential $V(\abs{x}) = -1/\abs{x}$,
  along with its modified version, $\wb{V}(\abs{x}) \equiv \chi(x)
  V(\abs{x}) + \paren{1-\chi(\abs{x})} V(L)$, which is consistent with
  \eqref{eq:vavg}. The modified potential $\wb{V}$ is identical to $V$ for
  $\abs{x} < L-\sigma$ to within the double machine precision, and is
  smoothly truncated to the constant value $V(L)$ for $\abs{x} > L$.}
  \label{fig:cutoff}
\end{figure}

We use a partition of unity to make the transition between $V[\rho](x,t)$
and $v(t)$ smooth, thereby avoiding numerical artifacts caused by
irregularities in the potential. 
Let $\chi: \RR \to \RR$ be a bump function, which is smooth and satisfies
\[\chi(x) =
\begin{cases}
  0 & \text{if } \abs{x}>L \\
  1 & \text{if } \abs{x}<L-\sigma
\end{cases}\]
for a parameter $\sigma > 0$. The function
\[\chi(x) = \frac12 \paren{\erf\paren{11.6 \, \frac{L-\sigma/2-x}{\sigma}} -
\erf\paren{11.6 \, \frac{-L+\sigma/2-x}{\sigma}}} \]
satisfies these properties to within the double machine precision. For $x \in
\RR^d$, we can define a radial bump function by $\chi(x) \gets
\chi\paren{\abs{x}}$
for $x \in \RR^d$. Then
\begin{equation} \label{eq:vmod}
  \wb{V}[\rho](x,t) = \chi(x) V[\rho](x,t) + \paren{1-\chi(x)} v(t)
\end{equation}
is equal to $V[\rho](x,t)$ for $\abs{x} < L-\sigma$, and $v(t)$ for
$\abs{x} > L$. The modified potential $\wb{V}[\rho]$ is therefore of the form required by the
method described in the previous sections, and we propose to solve
\eqref{eq:tdkse} with $V$ replaced by $\wb{V}$. An example of this
truncation procedure, for $V$ taken to be the standard Coulomb
potential, is shown in Figure \ref{fig:cutoff}. We note that anisotropic
truncations of the potential could also be considered, and may be useful, for
example, in systems with strongly polarized potentials.

\subsection{Summary of the method}

In Section \ref{sec:fcd}, we noted that in order to apply our method,
a potential equal to a
constant $v(t)$ outside of $[-L,L]^d$ should first be reduced to a
compactly supported potential by the gauge transformation $V[\rho](x,t) \gets
V[\rho](x,t) - v(t)$. Replacing $V$ with $\wb{V}$ in \eqref{eq:tdkse2}, making this gauge
transformation, and noting that the corresponding phase shift in
$\psi_j$ does not affect $\rho$, we obtain the following set of
equations:
\begin{equation} \label{eq:tdksemod}
  \begin{aligned}
    i \partial_t \psijtil(x,t) &= \paren{-\frac12 \nabla^2 + i
    \vecA(t) \cdot \nabla + \chi(x) \paren{V[\rho](x,t)-v(t)}} \psijtil(x,t) \\
    \psijtil(x,0) &= \psijztil(x) \\
    v(t) &= \frac{\Gamma\paren{d/2}}{2 \pi^{d/2} L^{d-1}}
    \int_{S_L^{d-1}} V[\rho](x,t) \, dS(x) \\
    \rho(x,t) &= \sum_{j=1}^N \abs{\psijtil(x,t)}^2 \\
    \psi_j(x,t) &= e^{-i \int_0^t v(s) \, ds} \psijtil(x,t).
  \end{aligned}
\end{equation}
Our proposal is to solve these equations by the method described in
Section \ref{sec:fcd}. In this way, by
making a simple modification of the potential, we have avoided the need
for artificial boundary conditions. We note that one
must keep track of the indefinite integral $\int_0^t v(s) \, ds$ in order to
recover $\psi_j$ from $\psijtil$. This can be done using a
quadrature rule of sufficiently high order to match the order of
accuracy of the time discretization.

\section{Numerical results: absorption and photoelectron spectra}
\label{sec:results}

We next present numerical results demonstrating the performance of the
FCD method for TDDFT applications which require a good description of 
electronic excitations to continuum states. More specifically,
we consider two experimentally important problems: (1) the calculation of the absorption cross section of a 
molecule in free space which, in the linear regime, is equivalent to the ionization 
cross section, and (2) the nonlinear problem of calculating the photoelectron spectrum 
of a molecule in the tunneling regime. We find that FCD allows us
to significantly decrease the computational domain size $L$, as compared
with a simulation using CAPs, while maintaining similar quality results.

The calculation of the absorption cross section in the linear regime
using real time TDDFT is a standard procedure~\cite{Marques.2011,DeGiovannini.2018}.
The system is excited by including a small
momentum shift in the initial condition of the Kohn-Sham orbitals,
$\psi_{j,0}(x) \gets e^{i \lambda \cdot x} \psi_{j,0}(x)$. The spectrum is then
given by
\[S_{\alpha \beta}(\omega) = \frac{4 \pi\omega}{\lambda_\beta} \Im \int_0^\infty e^{i \omega t}
\paren{D_\alpha(t)-D_\alpha(0)} \, dt,\]
where 
$D(t)=\int_{\RR^d} x \rho(x,t) \, dx$ is the dipole moment of the
electron density, and $1 \leq \alpha,\beta \leq d$. In all calculations below, we
take $\lambda = \lambda_1 e_1$, with $e_1$ the unit vector aligned with
the $x_1$ axis, and measure $S_{11}(\omega)$. We therefore make the
replacement $\lambda \gets \lambda_1$ and $S(\omega) \gets
S_{11}(\omega)$ for simplicity.
We approximate $S$ by integrating only over a finite
propagation time.
The spectrum contains peaks corresponding to 
the allowed energy transitions of the system associated with the 
absorption of a photon of frequency $\omega$.

The photoelectron spectrum can be calculated using
tSURFF~\cite{Wopperer.2017}, a technique which has been implemented in
combination with
TDDFT~\cite{Sato.2018rl8,Krecinic.2018,Trabattoni.2020}. 
The method obtains the spectrum by accumulating the momentum-projected flux of the 
photoionization current through a ``flux surface,'' which we take to be
a sphere $S_R$ of radius
$R$ enclosing the system
of interest. More specifically, we first compute the momentum-resolved spectrum
$P(k) =\sum_{l=1}^N \left| \int_0^\infty dt\,
\int_{S_R} dS(x) \cdot j_{k,l}(x,t) \right|^2$, which
represents the probability of measuring electrons of a given momentum
vector $k$ escaping from the flux surface. Here $j_{k,l}(x,t)$ is the current
density of the photoelectron corresponding to the $l^\text{th}$ Kohn-Sham state, projected
onto Volkov waves, which are the solutions of the free particle TDSE
with an applied field $A(t)$ \cite[Eqs. 12, 20]{DeGiovannini.2018cms}.
In one spatial dimension, $\int_{S_R} dS(x)$ becomes a sum over the two points $x =
\pm R$.
From $P(k)$, we compute the energy-resolved spectrum, or photoelectron spectrum,  
by angular integration as $P(E) = \int_{|k|^2 = 2 E} P(k) \, dS(k)$,
which corresponds to the outgoing electron kinetic energy $E = |k|^2/2$.
$P(E)$, which represents the probability of measuring an electron with
kinetic energy $E$,
is typically used to interpret the ionization process in terms of energy conservation between 
the electrons and the field. 
We refer the reader to  \cite[Sec. 3]{DeGiovannini.2018cms} for further
details on the derivation of the photoelectron spectrum using tSURFF.

In simulations of both absorption and photoelectron spectroscopy, the observables can
be calculated with high accuracy using only the part of the wavefunction
contained in a bounded computational domain, provided that it remains
free of spurious boundary reflections.

We performed all numerical experiments using the Octopus code~\cite{Tancogne-Dejean.2020}, 
with an implementation of the FCD method.
For all FCD calculations, we use the time stepping method described in Appendix
\ref{app:propagator} with order $p=8$, and the truncation parameter
$\sigma = 0.03L$. We observe that the performance is fairly insensitive
to the specific choice of $\sigma$. The height of the contour $\Gamma$
is chosen according to the guidelines described in Ref.
\cite{kaye22_tdse} with an error tolerance $\varepsilon = 10^{-8}$.
For all calculations using CAPs, we use the enforced time-reversal
symmetry time stepping scheme described in Ref.~\cite{gomezpueyo18}, and the CAP defined in \cite[Eq.
17]{DeGiovannini.2015} with strength parameter $\eta =-0.2$, which is chosen in order 
to have a wide window of energy absorption \cite[Fig.~5]{DeGiovannini.2015}. 
The CAP width $w$ must be adjusted to provide sufficient absorption
without requiring too large of a computational domain. We
emphasize that the FCD approach does not require tuning parameters like
$\eta$ and $w$, so that in practice, it eliminates a costly scan over parameter
space, as is required by CAPs.

In the simulations of photoemission spectroscopy, we use a vector
potential of the form
\begin{equation} \label{eq:pulse}
  A(t) = A_0 \sin^2\paren{\pi t/T} \cos(\omega_0 t) \theta(T-t) \, e_1,
\end{equation}
which represents a laser pulse linearly polarized along the first coordinate
direction. Here $T$ is the duration of the pulse, $A_0$ is the pulse
strength,
$\theta$ is the Heaviside function, $\omega_0$ is the carrier angular frequency, 
and $e_1$ is the unit vector aligned with the $x_1$ axis (or $e_1 = 1$
for $d = 1$). 
The pulse strength is given by $A_0 = \sqrt{0.28\times 10^{-16} I c} \,
/\omega_0$, where $I$ is the laser peak intensity in 
units of W/cm$^2$ (converted here to a.u.), and $c$ is the speed of light.

All calculations are converged with respect
to the grid spacing $\Delta x$ and the time step $\Delta t$, and with
respect to the grid spacing along the contour $\Gamma$ for FCD
calculations. In this way, we
focus solely on convergence with respect to the computational domain size $L$.

\subsection{One-dimensional LiH dimer} 
\label{ssub:lih}
We first consider a one-dimensional model of LiH, a simple system 
with a nontrivial multi-electron
structure containing both core and weakly-bound valence states. 
A majority of molecules in three dimensions present a similar energy level structure.
The effective 1D Hamiltonian for the dimer is determined by an ionic potential defined 
as
\begin{equation}
  v_{\rm ion}(x)=-\sum_{a=1}^2 \frac{Z_a}{\sqrt{(x- X_a)^2+\alpha}}.
\end{equation}
Here the nuclear charge $Z_a$ is equal to 3 for Li and 1 for H, and
$X_{1,2}=\pm 1.15$ are the centers of the two nuclei. Each is described by a
soft-core potential with regularization parameter $\alpha=0.5$ a.u.$^2$~\cite{Sato.2013}. 
We simulate $N=4$ interacting electrons, arranged in two doubly occupied
orbitals, using the one-dimensional
local-density approximation (LDA) for $v_{\rm
xc}$~\cite{Helbig.2011,Casula.2006}.

\begin{figure}
  \centering
  \includegraphics[width=0.9\textwidth]{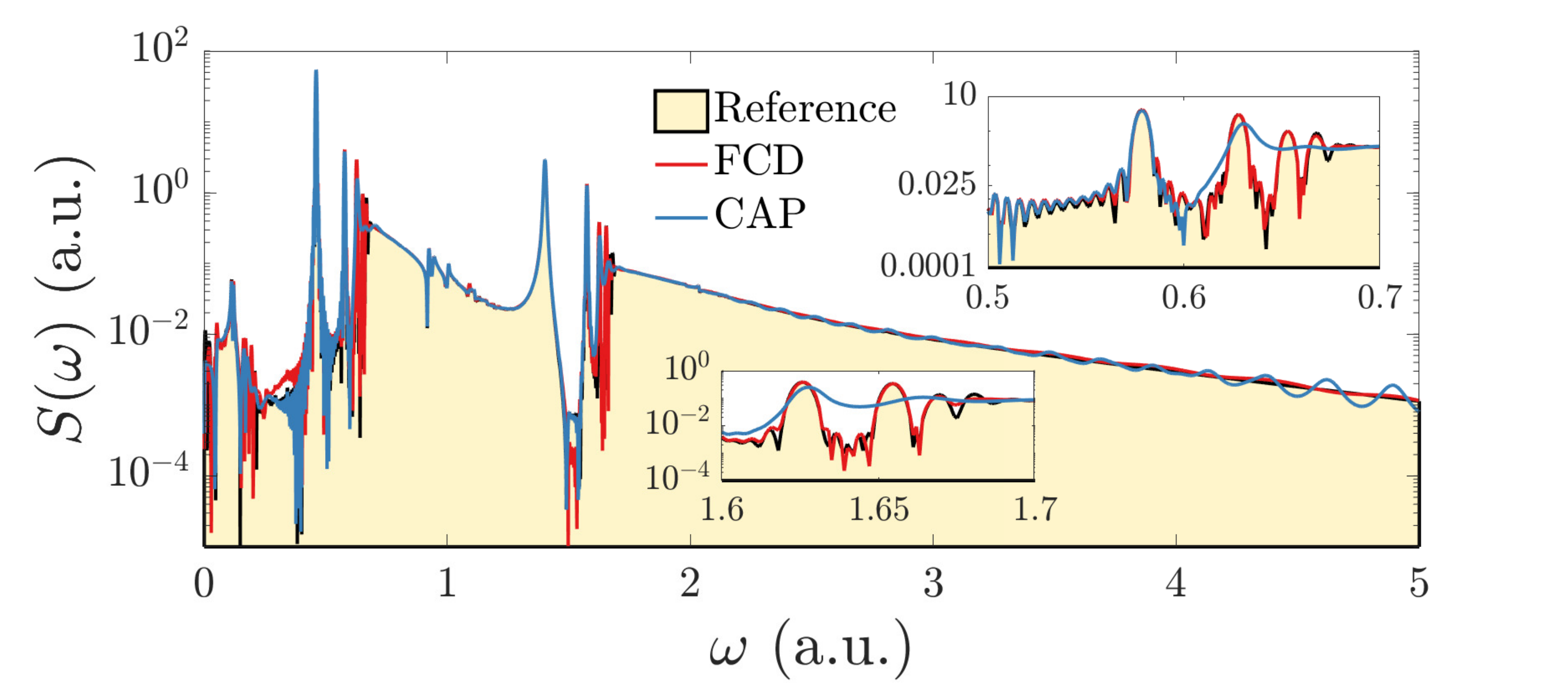}
  \caption{\label{fig:lih_abs}
  Computed absorption spectrum for the one-dimensional LiH example. We show results
  from a well-converged reference on a large domain, along with FCD, and a CAP of
  width $w = 10$~a.u., both on domains of width
  $L = 30$~a.u.. 
  The insets are zooms on the first and second continuum thresholds.}
\end{figure}

In Fig.~\ref{fig:lih_abs} we compare the absorption cross section
computed with FCD, and with a CAP, on computational domains of the same
size. We propagate for approximately $24$ fs, and use a momentum
parameter $\lambda = 0.01$ a.u. to excite the initial state. In black, we show a reference
spectrum, obtained using a CAP, which was converged with respect to all simulation
parameters, including the computational domain size $L$ and the CAP width
$w$. Since the system has two well-separated bound states, the cross section
shows two distinct ionization thresholds near $0.7$~a.u. and $1.7$~a.u., corresponding to the ionization
energies of the two states.
A calculation using FCD on a domain of width $L = 30$ a.u. is largely
successful in capturing the detailed structure of the spectrum. On the
other hand, the result using a CAP of width $w = 10$ a.u. on a domain of
width $L = 30$ a.u. is poorer, exhibiting spurious oscillations at high
energies, and a discrepancy localized near
the peaks corresponding to the ionization thresholds.
Here $w$ was varied to achieve the best possible result,
and it cannot be further improved as long as $L$ is kept fixed. In our
experiments, achieving a similar quality result with a CAP required a
domain of width at least $L = 60$, using a CAP of width $w = 20$. For both
the FCD and CAP simulations, we used a converged grid spacing $\Delta x = 0.3$ a.u.
and time step $\Delta t \approx 0.001$ fs.

\begin{figure}
  \centering
  \includegraphics[width=0.9\textwidth]{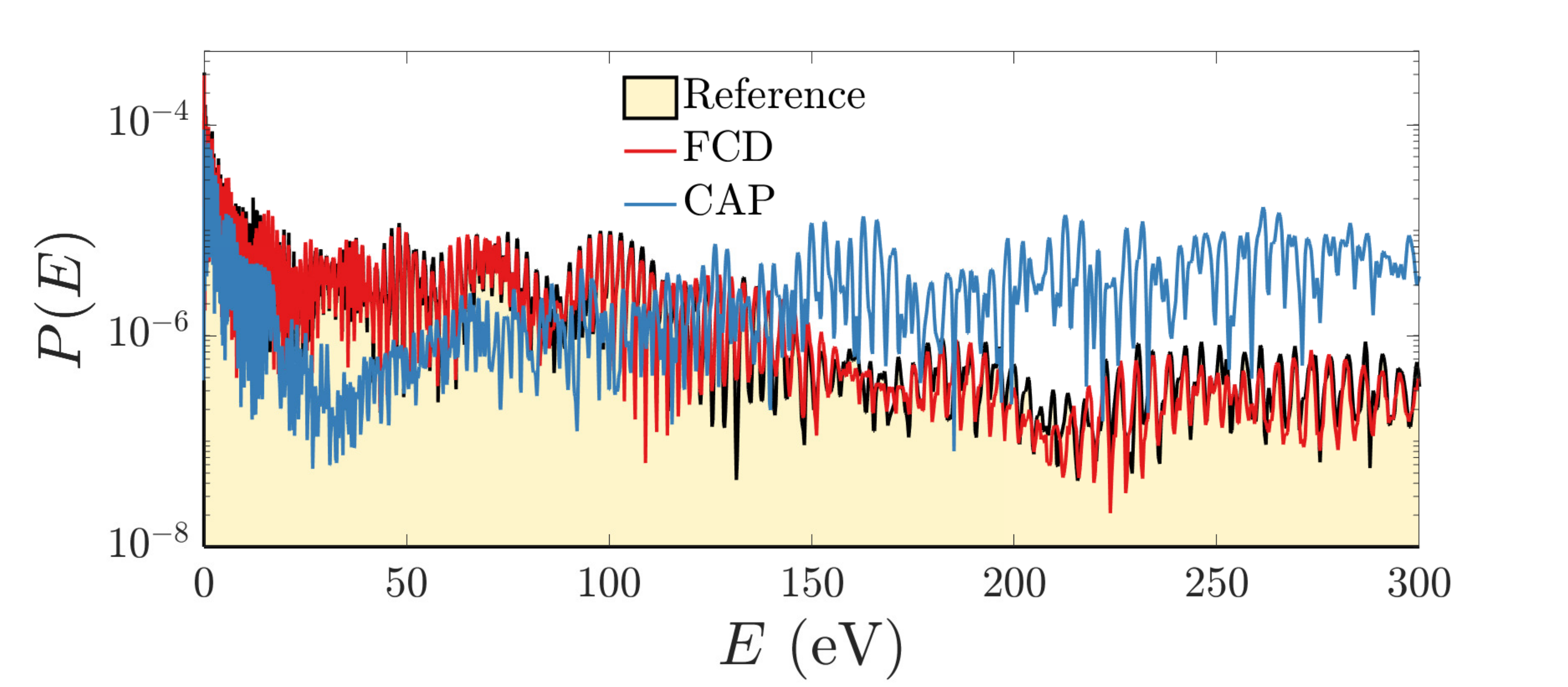}
  \caption{\label{fig:Lih_pes}
  Computed photoemission spectrum for the one-dimensional LiH example. We show results
  from a well-converged reference on a large domain, along with FCD, and a CAP of
  width $w = 10$~a.u., both on domains of width $L = 50$~a.u..}
\end{figure}

We next investigate the nonlinear effects of strong field ionization. 
We apply a $60$ fs pulse of the form \eqref{eq:pulse} with intensity $I=7.7 \times 10^{13}$~W/cm$^2$ 
and frequency $\omega_0=0.954$~eV, which is sufficiently strong to turn an
ionized electron back toward its parent cation
once the electrons have tunneled into the continuum.
We estimate the quiver radius of this field (the oscillation
amplitude of a free electron subjected to the field) as $38$ a.u..
We propagate the wavefunction for $90$ fs, and place the flux surface
at a radius $R = 40$ a.u. from the origin.
Photoelectrons emerging from direct and rescattering trajectories
interfere to form the photoelectron spectrum depicted in Fig.~\ref{fig:Lih_pes}. 
The spectrum presents characteristic features of strong field dynamics, 
such as oscillations mirroring the pulse frequency $\omega_0$,
which is typical of multiphoton processes. 
The decay behavior of the spectrum
reflects the classical approximation of rescattering dynamics. It
predicts a maximum energy of electron ejection at approximately $10
U_p$, where $U_p=A_0^2/4$ is the ponderomotive energy~\cite{Lin.2018}. 
In this case, we have $U_p \approx 12$
eV, and observe an onset of decay in the spectrum near the expected
energy. In principle, the spectrum should decay to zero from this point,
but in practice the finite time of simulation yields a finite tail~\cite{Morales.2016}. 

In this case, using FCD on a domain of width $L = 50$ is sufficient to
obtain good agreement with the converged reference. Using the same
domain size, we then take the largest
CAP width possible, $w = 10$, so that the absorbing region remains
outside of the flux surface. The quality of
the result is poor across all energies. 
Increasing the domain width to $L = 120$ and using a CAP of width $w = 40$
yields a result of comparable quality to the FCD calculation. Here, we
used the same converged grid spacing and time step size as in the
absorption spectroscopy simulation.

\subsection{Two-dimensional benzene molecule}
\label{ssub:benzene}
We next consider a model of a benzene molecule 
given by a hexagonal arrangement of six soft-core atoms in dimension $d
= 2$~\cite{Ceccherini.2001o6p,Castiglia.2015}:
\begin{equation}
  v_{\rm ion}(x)=-\sum_{a=1}^6
  \frac{1}{\sqrt{|x-X_a|^2+\alpha}}.
\end{equation}
We use the regularization parameter $\alpha=0.45$ a.u.$^2$, and atomic centers
$X_a=\big(R\cos(a\pi/3),R\sin(a \pi/3)\big)$
with $R=2.63$~a.u..
We simulate an interacting system of $N = 6$ electrons in three
doubly-occupied orbitals, using two-dimensional LDA \cite{Attaccalite.2002}.
\begin{figure}
  \centering
  \includegraphics[width=0.9\textwidth]{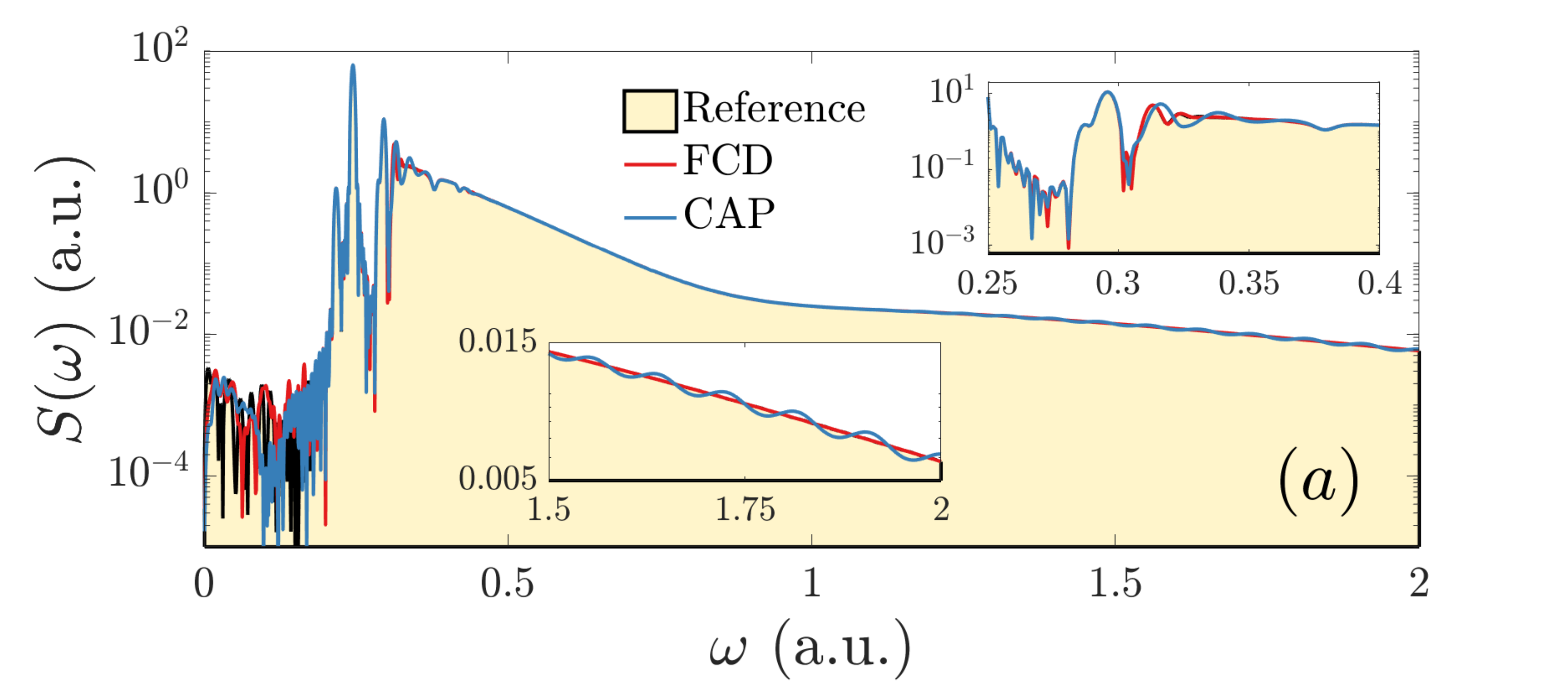}
  \caption{\label{fig:benzene_abs}
  Computed absorption spectrum for the two-dimensional benzene example. We show results
  from a well-converged reference on a large domain, along with FCD, and a CAP of
  width $w = 30$~a.u., both on domains of size $L = 60$~a.u..
  The insets are zooms on the continuum threshold and the high energy end of the spectrum.}
\end{figure}
The system contains a doubly degenerate HOMO state and a relatively
close HOMO-1 state, as reflected by the single clean ionization
threshold appearing near 0.46 a.u. in the absorption cross section shown
in Fig.~\ref{fig:benzene_abs}. The first ionization threshold is overshadowed
by bound-to-bound transitions.

Fig. \ref{fig:benzene_abs} compares the absorption cross section obtained
using a converged reference calculation with that using FCD with a
domain of size $L = 60$, and a CAP of width $w = 30$ on a domain of
the same size ($w$ was varied to achieve the best possible result). We
used a propagation time of approximately $24$ fs, and a momentum
parameter $\lambda = 0.01$ a.u. to excite the initial state. 
As in the one-dimensional example, the discrepancy is
most noticeable in the large energy regime and near the ionization
threshold. We note that although the spectrum should decay to zero at
low energies, the finite time of propagation yields some non-zero
residual, and therefore the reference calculation should not be
considered to be accurate below $\omega \approx 0.2$ a.u..
A domain of size $L = 140$ with a CAP
of width $w = 60$ was required to obtain a result of similar quality to
the FCD calculation. For both
the FCD and CAP simulations, we used a converged grid spacing $\Delta x = 0.4$ a.u.
and time step $\Delta t \approx 0.001$ fs.

\begin{figure*}[t]
  \centering
  \includegraphics[width=0.9\textwidth]{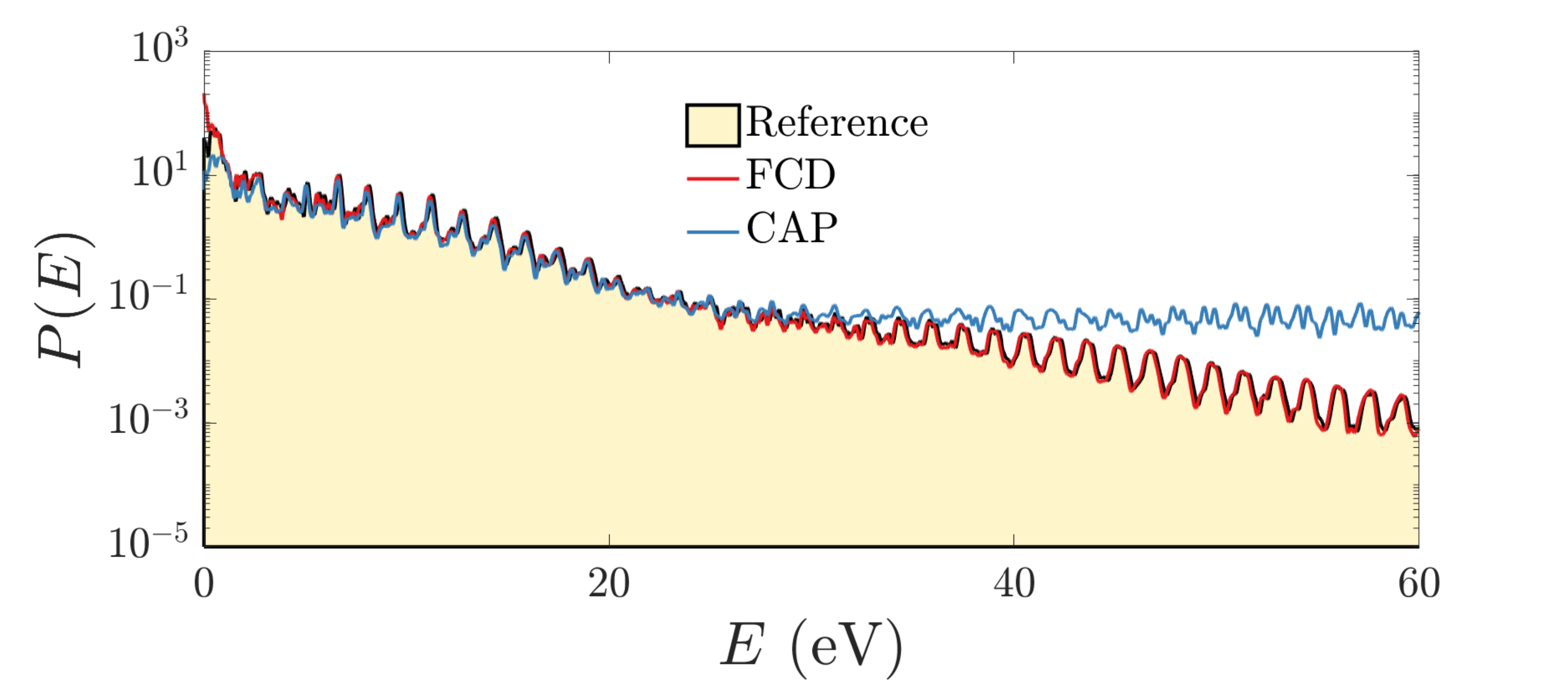}
  \caption{\label{fig:benzene_pes}
  Angle-integrated photoemission spectrum for the two-dimensional benzene example. We show results
  from a well-converged reference on a large domain, along with FCD, and a CAP of
  width $w = 10$~a.u., both on domains of size $L = 50$~a.u..}
\end{figure*}
Fig. \ref{fig:benzene_pes} shows the photoelectron
spectrum, obtained using a $64$ fs laser pulse
with $I=10^{14}$~W/cm$^2$ and $\omega_0=1.55$ eV. We place the flux surface
at a radius $R = 40$ from the origin. We use a domain of size
$L = 50$ for FCD, as well as a CAP of width $w = 10$ on a domain
of the same size. Here, the CAP is taken to be as wide as
possible while still remaining outside of the flux surface. The CAP produces a significant discrepancy with the reference
calculation in the large energy regime.
A domain of size $L = 80$ with a CAP of width $w = 40$ was required
to obtain a result of similar quality to the FCD calculation. For both
the FCD and CAP simulations, we used a converged grid spacing $\Delta x = 0.4$ a.u.
and a time step $\Delta t \approx 0.002$ fs.

An important observable which can be obtained from photoelectron
experiments is the photoelectron angular distribution (PAD), which
depicts the probability of measuring an electron ionized with a given momentum
as a two dimensional map $P(k_x,k_y)$. 
PADs obtained by strong field ionization are commonly used to study the
relationship between the electronic structure of a system and its
dynamics under a strong perturbation.
\begin{figure}
  \centering
  \includegraphics[width=0.48\textwidth]{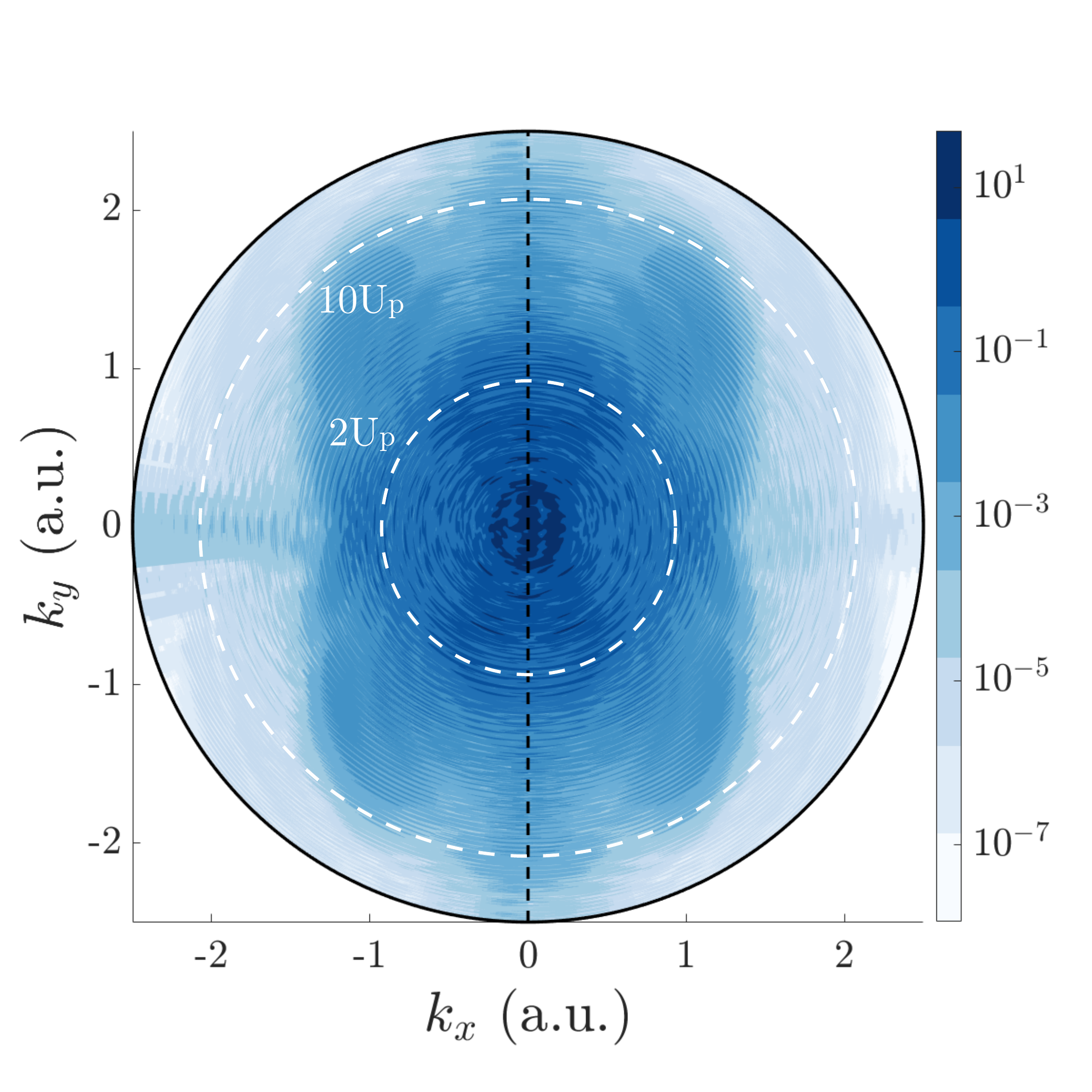}
  \includegraphics[width=0.48\textwidth]{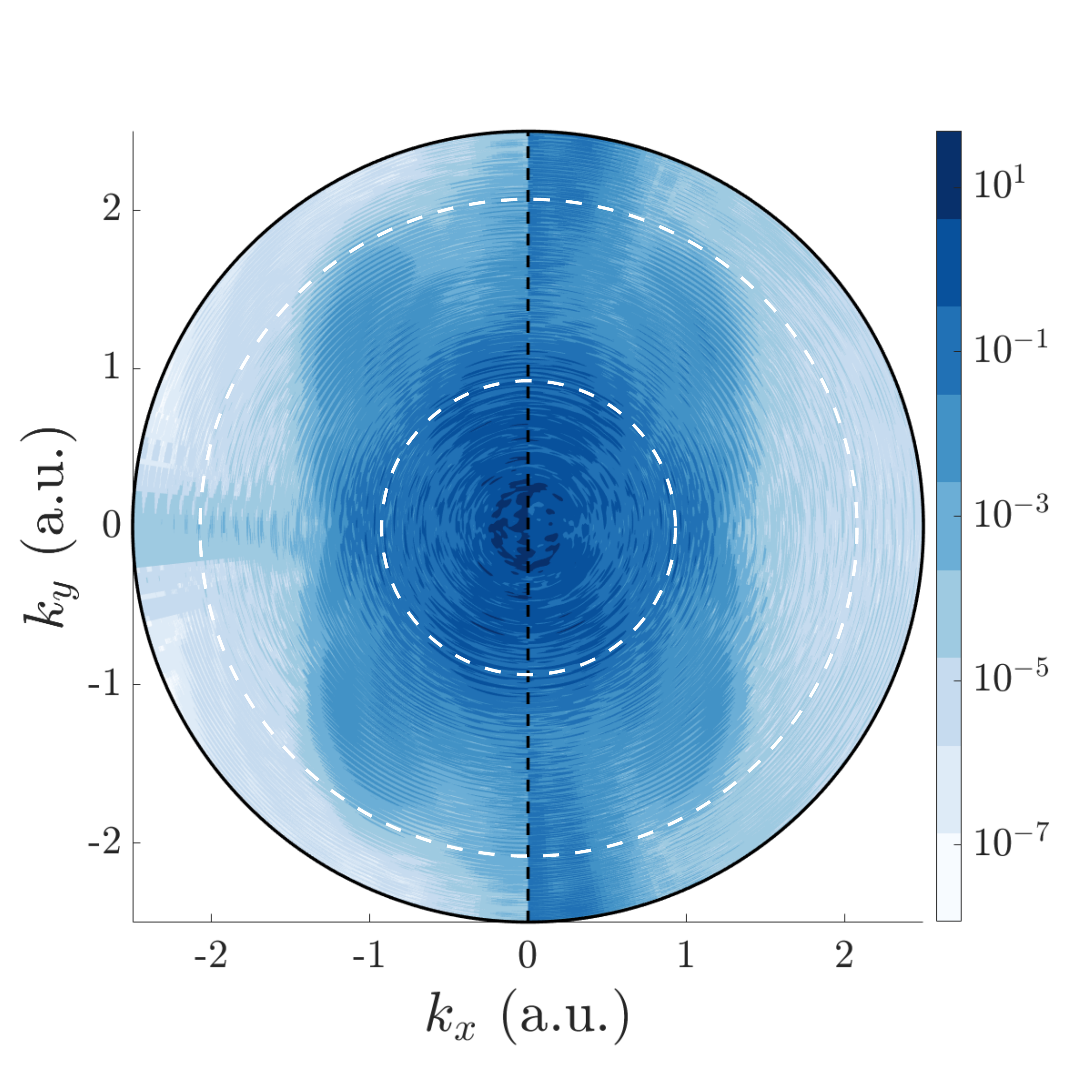}
  \caption{\label{fig:benzene_pad}
  Photoelectron angular distribution for two-dimensional benzene
  example. Results for a well-converged reference are given in the left
  half of each plot. The FCD result on a domain of size $L = 50$
  a.u. is given on the right half of the
  first plot, and the CAP result on a domain of the same size, and a CAP
  width $w = 10$~a.u., is given on the right half of the second plot.}
\end{figure}
Fig. \ref{fig:benzene_pad} shows the PAD corresponding to the spectrum 
given in Fig. \ref{fig:benzene_pes}. 
It is known that photoelectrons ionized with energies
below $2U_p$ are mostly ejected via multiphoton absorption in a process called above threshold 
ionization, and those with energies between $2U_p$ and $10U_p$
through rescattering processes involving a boost in kinetic energy
obtained from the applied field.
Considering this energy separation, several important properties of the system and its dynamics 
can be obtained.
For instance, in the rescattering region, an intricate picture is formed by the superposition of 
rings centered at the value of the field at the rescattering time. From
this analysis, one can infer structural properties of a system, such as its bond lengths~\cite{Lin.2018}. 
Similarly, by analysing the ``spider-leg'' features formed by interference between direct and rescattered 
electrons, one can recover information on the time between rescattering events~\cite{Huismans.2011}.
A proper understanding of the details of the PAD, especially for large and complex molecules,
is an active and significant area of research~\cite{Wolter.2015, Amini.2020, DeGiovannini.2022}. 

The comparison in Fig. \ref{fig:benzene_pad} between the PAD obtained with FCD and with a CAP
provides a much more detailed picture than Fig.~\ref{fig:benzene_pes}.
The main discrepancy between the reference and the CAP is concentrated along the 
laser polarization axis, $k_y$, where most of the rescattering dynamics
take place.
The FCD calculation, on the other hand, is in good agreement with
the reference throughout the momentum domain. The low-energy regime (below
$\omega \approx 1.5$ eV in Fig. \ref{fig:benzene_pes} and
$\abs{k} \approx 0.33$~a.u. in Fig. \ref{fig:benzene_pad}) corresponds to
photoelectrons ejected with the smallest momentum, which suffer the
greatest distortion from the truncation of the Coulomb tail
\cite{Arbo.2008}. Nevertheless, the FCD method reproduces the correct
qualitative features of the PAD in this region much more effectively
than the CAP method.

\section{Conclusion} \label{sec:conclusion}

We have demonstrated an implementation of the newly-introduced FCD method for TDDFT
simulations in chemistry and materials science which handles the problem of spurious boundary reflections
for free space problems in a different manner than standard methods
like CAPs, mask functions, PMLs, and ECS. In particular, the underlying
TDKSE is modified only by a smooth, controlled truncation of long range
potentials, and the resulting problem is solved in a mathematically
exact manner. To understand the difference with previous methods, one
can consider a problem with a compactly supported potential. Whereas the methods
mentioned above must still suppress boundary reflections using an
absorbing boundary layer, introducing some domain truncation error, the
FCD method has no such truncation error in this setting. Our approach is therefore to first reduce the TDKSE to this case, and then
apply the FCD method. We find that for simulations of absorption and
photoelectron spectroscopy, this yields high quality spectra
with significantly smaller computational domain sizes than CAPs.

In realistic, three-dimensional TDDFT simulations using typical
artificial boundary schemes, unaffordably large domains are often
required to avoid excessive pollution of results by spurious boundary reflections. The FCD
approach represents a promising opportunity for improvement in this
respect. A scalable parallel implementation of the method in three
dimensions is in progress, as well as an extension to
periodic solids with one free dimension.


\acknowledgments

We acknowledge financial support from the European Research Council
(ERC-2015-AdG-694097), the Cluster of Excellence
Advanced Imaging of Matter (AIM), Grupos Consolidados (IT1453-22), 
SFB925, and the Max Planck--New York City Center
for Non-Equilibrium Quantum Phenomena. The Flatiron Institute is a division of the
Simons Foundation.

\appendix

\section{High-order time propagation, and other implementation details} \label{app:propagator}

We suggest a particular method of high-order accurate time propagation,
and leave comparison with other possible methods as an important direction of future
research. The state index $j$ of $\psi_j$ is again made explicit in this
section for clarity. We first
rewrite \eqref{eq:tdsegamma} in integral form,
\begin{equation} \label{eq:psihatinteq}
  \psijhat(\zeta,t) = e^{- \frac{i}{2} \zeta\cdot\zeta t + i \zeta \cdot \varphi(t)}
  \psijzhat(\zeta) - i \int_0^t e^{-\frac{i}{2} \zeta\cdot\zeta (t-s) 
    + i \zeta \cdot (\varphi(t) - \varphi(s))} \Vpsijhat(\zeta,s) \, ds,
 \end{equation}
with
\[\varphi(t) \equiv \int_0^t \vecA(s) \, ds.\]
This is simply the variation of parameters solution, treated as an ordinary differential equation for
each fixed $\zeta$, with $\Vpsijhat$ considered as an inhomogeneity.
Introducing a discrete time step $\Delta t$, simple
algebraic manipulations allow us to eliminate the full memory dependence
in \eqref{eq:psihatinteq} by writing it as a recurrence in time:
\begin{equation} \label{eq:psihatrecur}
  \psijhat(\zeta,t) = e^{-\frac{i}{2} \zeta\cdot\zeta \Delta t + i \zeta \cdot
      \paren{\varphi(t) - \varphi(t-\Delta t)}} \psijhat(\zeta,t-\Delta t)
    - i \int_{t - \Delta t}^t e^{-\frac{i}{2} \zeta\cdot\zeta (t-s) + i \zeta \cdot
      (\varphi(t) - \varphi(s))} \Vpsijhat(\zeta,s) \, ds.
\end{equation}
Thus, at each time step, we modify the solution by a damping phase
factor and a local update integral.

This update integral can be discretized to high-order accuracy by the
implicit Adams-Moulton multistep method. In this approach, to achieve
$p^\text{th}$ order accuracy,
the integrand is replaced by the polynomial interpolant of its values at
$s = t, t-\Delta t, \ldots, t - (p-1) \Delta t$, and the resulting
integrals are computed analytically. This yields an expression
\begin{equation} \label{eq:amsemidisc}
  \begin{multlined}
    \psijhat(\zeta,t) + i \mu_0 \Delta t \Vpsijhat(\zeta,t) \approx e^{-\frac{i}{2}
    \zeta\cdot\zeta \Delta t + i \zeta \cdot \paren{\varphi(t) -
    \varphi(t-\Delta t)}} \psijhat(\zeta,t-\Delta t) \\
    - i \Delta t \sum_{k=1}^{p-1} \mu_k e^{-\frac{i}{2} \zeta\cdot\zeta k \Delta t
    + i \zeta \cdot (\varphi(t) - \varphi(t-k\Delta t))}
    \Vpsijhat(\zeta,t-k\Delta t),
  \end{multlined}
\end{equation}
with $\mu_j$ the Adams-Moulton weights. For a procedure
to obtain the weights, along with tabulated weights for
methods of up to eighth order, we refer the reader to Ref.
\cite[Chap. 24]{butcher16}. At time $t$, the right hand side is known,
and for brevity we write it as
\begin{equation} \label{eq:frhs}
\wh{f_j}\paren{\zeta,t-\Delta t} \equiv e^{-\frac{i}{2} \zeta\cdot\zeta \Delta t + i \zeta \cdot
  \paren{\varphi(t) - \varphi(t-\Delta t)}} \psijhat(\zeta,t-\Delta t) \\ -
  i \Delta t \sum_{k=1}^{p-1} \mu_k e^{-\frac{i}{2} \zeta\cdot\zeta k \Delta t + i
  \zeta \cdot (\varphi(t) - \varphi(t-k\Delta t))}
  \Vpsijhat(\zeta,t-k\Delta t).
\end{equation}
Taking the deformed inverse Fourier transform \eqref{eq:psizift} then
gives
\[\paren{1 + i \mu_0 \Delta t V[\rho](x,t)} \psi_j(x,t) \approx
\frac{1}{(2\pi)^d} \int_{\Gamma^d} e^{i \zeta \cdot x}
\wh{f_j}\paren{\zeta,t-\Delta t} \, d\zeta \equiv f_j(x,t-\Delta t),\]
which can be
solved for $\psi_j$ by pointwise division to obtain
\begin{equation} \label{eq:nonlinstep}
  \psi_j(x,t) \approx f_j(x,t-\Delta t) / \paren{1 + i \mu_0 \Delta t
  V[\rho](x,t)}.
\end{equation}
This should be contrasted with a
typical implicit time stepping method, in which taking a time step
in general requires solving a linear system. It is a consequence of
using the integral formulation \eqref{eq:psihatinteq} that the system is diagonal
in this case. We have in \eqref{eq:nonlinstep} a nonlinear system of $N$ equations,
$j=1,\ldots,N$, coupled
via the density $\rho$ through the potential $V$. 


A good guess for the iteration can be obtained from a $p^\text{th}$ order explicit
Adams-Bashforth discretization of the update integral in \eqref{eq:psihatrecur}, obtained by
replacing the integrand with the polynomial interpolant of its values at
$s = t-\Delta t, \ldots, t - p \Delta t$. Following the same steps as
for the implicit Adams-Moulton discretization, we obtain the Adams-Bashforth guess
\begin{equation} \label{eq:absemidisc}
  \begin{multlined}
    \psijtil(x,t) \approx \frac{1}{(2\pi)^d} \int_{\Gamma^d} e^{i \zeta
    \cdot x} \wh{g_j}(\zeta,t-\Delta t) \, d\zeta \equiv g_j(x,t-\Delta
    t),
  \end{multlined}
\end{equation}
with
\begin{equation} \label{eq:grhs}
  \wh{g_j}(\zeta,t-\Delta t) = e^{-\frac{i}{2}
    \zeta\cdot\zeta \Delta t + i \zeta \cdot \paren{\varphi(t) -
    \varphi(t-\Delta t)}} \psijhat(\zeta,t-\Delta t)
    - i \Delta t \sum_{k=1}^p \nu_k e^{-\frac{i}{2} \zeta\cdot\zeta k \Delta t
    + i \zeta \cdot (\varphi(t) - \varphi(t-k\Delta t))}
    \Vpsijhat(\zeta,t-k\Delta t),
\end{equation}
for $\nu_k$ the Adams-Bashforth weights \cite[Chap. 24]{butcher16}.

We summarize the full numerical procedure as follows:
\begin{enumerate}
  \item Given $\psijhat(\zeta,t-\Delta t)$ and
    $\Vpsijhat(\zeta,t-k \Delta t)$ for $k
    = 1,\ldots,p$, compute $\wh{f_j}(\zeta,t-\Delta t)$ and
    $\wh{g_j}(\zeta,t-\Delta t)$
    using \eqref{eq:frhs} and \eqref{eq:grhs}, respectively, and
    compute $f_j(x,t-\Delta t)$ and $g_j(x,t-\Delta t)$ using the deformed
    inverse Fourier transform.
  \item Compute $\rho$ using the $\psijtil$ obtained from the Adams-Bashforth guess
    \eqref{eq:absemidisc}, and use it to compute an initial guess of
    $V[\rho]$. 
  \item Solve the nonlinear system \eqref{eq:nonlinstep} by
    self-consistent iteration; for a simple fixed point iteration, we
    would compute each $\psi_j$ from \eqref{eq:nonlinstep}, compute an
    updated value of $\rho$ and of $V[\rho]$, and repeat.
  \item Use the solutions $\psi_j$ to compute $\Vpsij(x,t)$ and
    take its Fourier transform to obtain $\Vpsijhat(\zeta,t)$. Then use
    \eqref{eq:amsemidisc} to compute $\psijhat(t)$, and advance to the
    next time step.
\end{enumerate}

We must mention two remaining issues to complete our description of the method.
First, we must accurately discretize and compute Fourier transforms
between $x \in [-L,L]^d$ and $\zeta \in \Gamma^d$. Ref.
\cite{kaye22_tdse} describes a high-order discretization scheme and an
FFT-based fast algorithm to compute these transforms efficiently. We
note that the algorithm for $d>1$ can be simplified considerably
compared with the prescription given there: given
an implementation of the algorithm for $d = 1$, one can simply use the separability of the complex-frequency
discrete Fourier transform to apply it dimension by dimension.

Second, we require an initialization procedure
to carry out the first $p-1$ time steps with $p^\text{th}$ order accuracy. For this,
we again refer to Ref. \cite{kaye22_tdse}, which describes a high-order Richardson
extrapolation method for initialization. Briefly, this method performs a
$p^\text{th}$ order time step by extrapolating the results of $j$
time steps of size $\Delta t / 2^j$, for $j =
0,\ldots,p/2-1$, obtained using the second order trapezoidal rule (the
Adams-Moulton method for $p=2$), which does not require initialization.

\bibliographystyle{ieeetr-reversed}
\bibliography{fcdtddft}

\end{document}